\documentclass[pdflatex,sn-mathphys-num]{sn-jnl}% Math and Physical Sciences Numbered Reference Style
\usepackage{graphicx}%
\usepackage{multirow}%
\usepackage{amsmath,amssymb,amsfonts}%
\usepackage{amsthm}%
\usepackage{mathrsfs}%
\usepackage[title]{appendix}%
\usepackage{xcolor}%
\usepackage{textcomp}%
\usepackage{manyfoot}%
\usepackage{booktabs}%
\usepackage{algorithm}%
\usepackage{algorithmicx}%
\usepackage{algpseudocode}%
\usepackage{listings}%
\usepackage{physics}
\usepackage{subfigure}

\theoremstyle{thmstyleone}%
\newtheorem{theorem}{Theorem}%  meant for continuous numbers
\newtheorem{proposition}[theorem]{Proposition}% 

\theoremstyle{thmstyletwo}%
\newtheorem{example}{Example}%
\newtheorem{remark}{Remark}%

\theoremstyle{thmstylethree}%
\newtheorem{definition}{Definition}%

\theoremstyle{thmstylethree}%
\newtheorem{corollary}{Corollary}%
\begin{document}

\title[Article Title]{Kernel-Preserving Dynamics and Symmetry Classification for Synchronization Subspaces}


\author*[1]{\fnm{Nicholas R.} \sur{Allgood}}\email{allgood1@umbc.edu}

\affil*[1]{\orgdiv{Department of Computer Science and Electrical Engineering}, \orgname{University of Maryland Baltimore County}, \orgaddress{\street{1000 Hilltop Circle}, \city{Baltimore}, \postcode{21250}, \state{MD}, \country{US}}}


\abstract{We study the preservation and stability of synchronization subspaces in tensor products of finite-dimensional Hilbert spaces. Given self-adjoint operators $T_A$ and $T_B$ on local subsystems, the synchronization subspace is defined as the kernel of the difference operator $K = T_A \otimes I - I \otimes T_B$. We establish two main results: First for $\epsilon$-compatible dynamics satisfying $||[H,K]|| \leq \epsilon$, we prove a sharp drift bound where any initially synchronized state deviates from the kernel at a  rate at most linear in time with slope $\epsilon$. We show by explicit construction that this estimate is optimal to leading order. Second in the presence of finite group symmetry, we show that the synchronization subspace coincides with the diagonal isotypic component in the tensor product decomposition and we characterize the algebra of synchronization-preserving dynamics as the intersection of the commutants of the group action and synchronization operator. }

\keywords{synchronization subspaces, kernel-preserving dynamics, $\epsilon$-compatible, quantum time transfer}


\pacs[MSC Classification]{81R15, 47A55, 20C35, 81P45}

\maketitle

\section{Introduction}\label{intro}

Let $H_A$ and $H_B$ be finite-dimensional Hilbert spaces equipped with self-adjoint operators $T_A$ and $T_B$ whose eigenvalues encode discrete time labels. The difference operator $K := T_A \otimes I - I \otimes T_B$ defines a synchronization subspace $\mathcal{K} := \operatorname{ker}(K)$ consisting of states for which the two subsystems yield identical time labels under measurement. This paper studies the preservation and stability of $\mathcal{K}$ under Hamiltonian dynamics, and classifies the algebraic structure of dynamics that preserve it. 

The question of when a distinguished subspace is preserved under approximately compatible dynamics has a long history in operator theory. Perturbative stability of invariant subspaces appears in the work of Davis and Kahan \cite{DavisKahan1970} on spectral perturbation bounds and approximate conservation laws arising from nearly commuting operators have been studied in the context of quantum lattice systems by Hastings and others \cite{Hastings2009, Lin1997}. The algebraic classification of subspace-preserving dynamics connects to the broader program of characterizing commutant algebras and symmetry-protected subspaces, with roots in the work of Kadison and Ringrose on operator algebras \cite{kadison_ringrose}. Our framework addresses these themes in a specific setting: kernel preservation for difference operators on tensor products that admits both sharp perturbative estimates and a complete representation-theoretic classification.

Our first main result (Theorem \ref{thm:driftbound}) establishes perturbative stability for $\epsilon$-compatible dynamics defined by the condition $||[H,K]|| \leq \epsilon$. We prove that for any initial state in $\operatorname{ker}(K)$ the drift satisfies $||K\ket{\psi(t)} || \leq \epsilon |t|$ and show by explicit construction that this bound is sharp to leading order. This provides a quantitative estimate on the degradation of synchronization under an approximately compatible evolution with the spectral gap of $K$ governing the fidelity of the decay rate. 

Our second main result (Theorem \ref{thm:synciso}) gives a representation-theoretic classification in the presence of a finite group symmetry $G$. When the observables (self-adjoint operators whose eigenvalues encode time labels) lie in the $G$-equivariant endomorphism algebras, the synchronization subspace coincides with the diagonal isotypic component $\oplus_\{\lambda \in \hat{G} \} V_\lambda \otimes V_\lambda$ in the tensor product decomposition. The algebra of synchronization-preserving dynamics is characterized as the intersection of the commutant of the group action with the commutant of $\mathcal{K}$, revealing synchronization as a structural invariant of the representation category rather than the property of individual operators. 

The framework is motivated by quantum time transfer, where distributed parties share entangled states and synchronize local clocks through correlated measurements \cite{PhysRevApplied.20.024064, lafler2024twoway}. In this setting, $T_A$ and $T_B$ are clock observables whose eigenvalues represent time labels, the kernel condition encodes perfect synchronization and $\epsilon$-compatible dynamics model the effect of experimental imperfections on timing correlations. Operational aspects of such protocols have been developed in several works, including entangled-photon and shared-oscillator implementations. The present paper provides a mathematical foundation for these protocols by identifying synchronization as an algebraic invariant and establishing quantitative stability guarantees.

The paper is organized as follows: Section \ref{setup} fixes notation and establishes exact preservation of the synchronization subspace as a baseline result. Section \ref{stability} develops the perturbative stability theory, including a sharp drift bound and fidelity estimates. Section \ref{representation} gives the representation-theoretic classification of synchronization-preserving dynamics under the finite group symmetry. Section \ref{discuss} discusses applications to quantum time transfer and outlines extensions to multipartite systems and categorical frameworks. 

\section{Setup and Notation}\label{setup}

$H_A$ and $H_B$ denote finite-dimensional Hilbert spaces and $H := H_A \otimes H_B$ denote their tensor product. All operators are bounded linear operators on finite-dimensional Hilbert spaces. A clock observable is a self-adjoint operator $T \in B(H)$ with discrete spectrum of $\sigma(T) = \{t_0, \cdots , t_{d-1} \} \subset \mathbb{R}$, whose eigenvalues we interpret as time labels. A self-adjoint Hamiltonian $H \in B(H)$ is compatible with $T$ if $[T,H] = 0$, in which case $T$ and $H$ are simultaneously diagonalizable and the eigenspaces of $T$ are preserved under the evolution $U(t) = e^{-iHt}$.

\begin{definition}
    Given the clock observables $T_A \in B(H_A)$ and $T_B \in B(H_B)$, we define the synchronization operator
    \begin{equation}
        K := T_A \otimes I - I \otimes T_B
    \end{equation}
    
    and the synchronization subspace
    \begin{equation}
        \mathcal{K} = \operatorname{ker}(K).
    \end{equation} 
\end{definition}

A state $\ket{\psi} \in \mathcal{K}$ satisfies $T_A \otimes I \ket{\psi} = I \otimes T_B \ket{\psi}$, meaning the two clock observables yield identical measurement statistics. When $T_A$ and $T_B$ have non-degenerate spectra with shared time labels, $\mathcal{K}$ is spanned by states of the form $\ket{j} \otimes \ket{j}$ where the eigenvalues match: $t_j^{(A)} = t_j^{(B)}$.

\begin{proposition}(Exact preservation) \label{prop:exactpres}

    Let $H_A \in B(H_A)$ and $H_B \in B(H_B)$ be self-adjoint with $[T_A, H_A] = 0$ and $[T_B, H_B] = 0$. Then the Hamiltonian $H:= H_A \otimes I + I \otimes H_B$ satisfies $[H,K] = 0$ and consequently $U(t)\mathcal{K} \subseteq \mathcal{K}$ for all $t \in \mathbb{R}$. 

    \begin{proof}
        Each term of $K$ commutes with each term of $H :[T_A \otimes I, H_A \otimes I] = [T_A, H_A] \otimes I = 0$ by hypothesis, $[T_A \otimes I , I \otimes H_B] = 0$ by disjoint tensor support and similarly for $I \otimes T_B$. Hence $[K, H] = 0$, and the invariance of $\operatorname{ker}(K)$ follows by functional calculus \cite{kadison_ringrose, reed_simon} $KU(t)\ket{\psi} = U(t)K\ket{\psi} = 0$ for any $\ket{\psi} \in \mathcal{K}$.
    \end{proof}
\end{proposition}

The set of all Hamiltonians compatible with a clock observable T forms a unital *-subalgebra of $B(H)$. When $T$ has a non-degenerate spectrum, this algebra consists of all operators diagonal in the eigenbasis of $T$. When $T$ has degenerate spectrum with spectral projections $\{P_\lambda\}$, the compatible algebra is isomorphic to $\displaystyle\bigoplus_\lambda B(\operatorname{ran}(P_\lambda))$ allowing unitary evolution within each eigenspace. 

Proposition \ref{prop:exactpres} characterizes exact preservation. The remainder of the paper addresses two questions: how does synchronization degrade when $[H,K]$ is small but non-zero and what algebraic structure governs in the presence of group symmetry?

\section{Perturbative Stability}\label{stability}

When $K$ commutes exactly with the Hamiltonian, the synchronization subspace is invariant. In practice, exact commutation is an idealization. We introduce a quantitative notion of approximate compatibility and establish sharp bounds on the resulting timing drift from synchronization. 

\begin{definition}($\epsilon$-compatible dynamics)
Let $T_A \in B(H_A)$ and $T_B \in B(H_B)$ be self-adjoint and define $K := T_A \otimes I - I \otimes T_B$. A self-adjoint Hamiltonian $H \in B(H_A \otimes H_B)$ is $\epsilon$-compatible with $K$ if $||[H,K]|| \leq \epsilon$.
\end{definition}

\begin{theorem}(Perturbative drift bound)\label{thm:driftbound}
Let $H$ be $\epsilon$-compatible with $K$ and let $\ket{\psi(0)} \in \operatorname{ker}(K)$ with $\ket{\psi(0)} = 1$. Then the evolved state $\ket{\psi(t)} = e^{-iHt}\ket{(0)}$ satisfies $\ket{K | \psi(t)} | \leq \epsilon|t|$.

\begin{proof}
    Define $\psi(t) := K\ket{\psi(t)}$. Then $d\psi / dt = -iKH\ket{\psi(t)} = -iHK\ket{\psi(t)} + i[H,K]\ket{\psi(t)}$ so $\psi$ satisfies a linear inhomogeneous equation with $\psi(0) = 0$. By Duhamel's formula \cite{Duahmel}, 
    \begin{equation}
        \psi(t) = -i \int_0^t e^{-iH(t-s)}[K,H] \ket{\psi(s)}ds
    \end{equation}

    Taking norms and using unitarity of $e^{-iH(t-s)}$ and $|\psi(s)| = 1:$  $|\psi(t)| \leq \int_0^t |[K,H]|ds \leq \epsilon | t|.$
\end{proof}
    
\end{theorem}

\begin{corollary}(Short-time stability)
For any $\delta > 0$, the evolved state satisfies $||K\ket{\psi(t)}|| \leq \delta $ for all $|t| \leq \delta / \epsilon$.

\begin{proof}
    Immediate from Theorem \ref{thm:driftbound}:$ ||K\ket{\psi(t)}|| \leq \epsilon|t| \leq \delta$ whenever $|t| \leq \delta / \epsilon$.
\end{proof}
\end{corollary}

\begin{corollary}(Fidelity decay).
Let $\kappa := \min\{|\lambda|: \lambda \in \sigma(K),\, \lambda \neq 0\}$ be the spectral gap of $K$. Then $\|\Pi_{\mathcal{K}}\ket{\psi(t)}\|^2 \geq 1 - \epsilon^2 t^2/\kappa^2$.
\begin{proof}
    Since $K$ acts on $\ker(K)^\perp$ with singular values at least $\kappa$, we have $\|\Pi_{\mathcal{K}^\perp} \ket{\psi(t)}\| \leq \|K\ket{\psi(t)}\| / \kappa \leq \epsilon|t|/\kappa$. The result follows from $\|\Pi_{\mathcal{K}}\ket{\psi(t)}\|^2 = 1 - \|\Pi_{\mathcal{K}^\perp} \ket{\psi(t)}\|^2$.
\end{proof}
\end{corollary}

\begin{example}(Sharpness of the drift bound)\label{ex:sharpdriftbound}

Let $H_A = H_B = \mathbb{C}^2$, $T_A = T_B = \sigma_z$ such that $K = \sigma_z \otimes I - I \otimes \sigma_z = \operatorname{diag}(0,2,-2,0)$. Take $H = (\epsilon / 2)(\ket{00}\bra{01} + \ket{01} \bra{00})$. A direct computation gives

\begin{equation}
  ||[H,K]|| = \epsilon
  \begin{pmatrix}
  0 & 1 \\
  -1 & 0 
\end{pmatrix}
\oplus 0_2,
\end{equation}
so $||[H,K]|| = \epsilon$. With the initial state $\ket{\psi(0)} = \ket{00} \in \operatorname{ker}(K)$, the evolved state satisfies $||K\ket{\psi(t)}|| = 2 | \sin (\epsilon t / 2 )| = \epsilon |t| + O(\epsilon^3 t^3)$. The leading-order drift saturates the bound of Theorem \ref{thm:driftbound}, so the estimate is optimal. No universal bound better than $\epsilon |t|$ holds over the class of all $\epsilon$-compatible Hamiltonians.

%Let $H_A = H_B = \mathbb{C}^2, T_A = T_B = \sigma_z$, so $K= \sigma_z \otimes I - I \otimes \sigma_z = \operatorname{diag}(0,2,-2,0)$. Take $H = (\epsilon / 2)(\ket{00}\bra{01} + \ket{01} \bra{00})$. Then $|[H,K]| = \epsilon$, and with the initial state $\ket{\psi(0)} = \ket{00} \in \operatorname{ker}(K)$, the evolved state gives

%\begin{equation}
 %   |K\ket{\psi(t)}| = 2|\sin(\epsilon t / 2)| = \epsilon |t| + O(\epsilon^3 t^3).
%\end{equation}

%The leading-order drift saturates the bound of Theorem \ref{thm:driftbound} showing the estimate is sharp.
    
\end{example}

\begin{remark}
    The exact drift in Example \ref{ex:sharpdriftbound} is periodic with period $2\pi\ / \epsilon$, so the true dynamics can return to perfect synchronization at finite times. The linear bound captures only the short-time behavior, which is the regime of physical interest and is optimal as a universal estimate. The bound is of Gr\"{o}nwall type \cite{Gronwall1919} and cannot be improved without additional structural assumptions on $H$. One natural source of such additional structure is symmetry. When a finite group $G$ acts on the system and the Hamiltonian is $G$-equivariant, the drift is confined to $G$-invariant subspaces, restricting the directions along which synchronization can degrade. 
\end{remark}

\section{Representation-Theoretic Classification}\label{representation}

We now classify synchronization-preserving dynamics in the presence of a finite group symmetry. The synchronization subspace is identified with the diagonal isotypic component in the tensor product decomposition, and the algebra of dynamics preserving it is characterized as the intersection of the commutant of the group action with the commutant of $K$.

Let $G$ be a finite group with unitary representations $\rho_A : G \rightarrow U(H_A)$ and $\rho_B : G \rightarrow U(H_B)$. Define the joint representation $\rho(g) := \rho_A(g) \otimes \rho_B(g)$ on $H = H_A \otimes H_B$. We then decompose each space into irreducibles:
\begin{align}
H_A = \oplus\{\lambda \in \hat{G}\} V_\lambda\otimes \mathbb{C}^{m_\lambda} \\
H_B = \oplus\{\lambda \in \hat{G}\} V_\lambda\otimes \mathbb{C}^{n_\lambda}
\end{align}

We then define the diagonal isotypic subspace:
\begin{equation}
    \mathcal{K}_G := \oplus\{\lambda \in \hat{G} \} V_\lambda \otimes V_\lambda
\end{equation}
\begin{theorem}(Synchronization as diagonal isotypic component) \label{thm:synciso}
    Let $T_A \in \operatorname{End}_G(H_A)$, $T_B \in \operatorname{End}_G(H_B)$ be self-adjoint and define $K := T_A \otimes I - I \otimes T_B$. Then:
    \begin{enumerate}
        \item $K$ commutes with $\rho(g)$ for all $g \in G$.
        \item If $T_A$ and $T_B$ assign the same scalar to each irreducible type, that is, $T_A \mid \{V_\lambda \} = a_\lambda I$ and $T_B \mid \{V_\lambda\} = a_\lambda I$ for all $\lambda$ then $\operatorname{ker}(K) \supseteq \mathcal{K}_G$. 
        \item If additionally $T_A, T_B \in Z(\mathbb{C}[G])$ and the central element separates representations that is the map $\lambda \mapsto \alpha_\lambda$ is injective on $\hat{G}$, then $\ker(K) = \mathcal{K}_G$.
    \end{enumerate}
\begin{proof}
    \textit{Part (i).} Since $T_A$ commutes with $\rho_A(g)$ and $T_B$ commutes with $\rho_B(g)$, we have 
    \begin{equation}
    [\rho(g), T_A \otimes I] = [\rho_A(g), T_A] \otimes \rho_B(g) = 0
    \end{equation} 
    and similarly $[\rho(g), I \otimes T_B] = 0$. Hence $[\rho(g), K] = 0$.

    \textit{Part (ii).} By Schur's lemma \cite{schur-duality}, $T_A$ and $T_B$ act as scalars $\alpha_\lambda$ and $\beta_\lambda$ on each $V_\lambda$. For any $v \otimes w \in V_\lambda \otimes V_\lambda$, we have $K(v \otimes w) = (\alpha_\lambda - \beta_\lambda)v \otimes w$. The condition $\alpha_\lambda = \beta_\lambda$ for all $\lambda$ gives $\mathcal{K}_G \subseteq \operatorname{ker}(K)$.

    \textit{Part (iii).} When $T_A, T_B \in Z(\mathbb{C}[G])$, they act as scalars $\alpha_\lambda$ on each $V_\lambda$ by Schur's lemma. For $v \otimes w \in V_\lambda \otimes V_\mu$ with $\lambda \neq \mu$, we have $K(v \otimes w) = (\alpha_\lambda - \alpha_\mu)v \otimes w$. Since the map $\lambda \mapsto \alpha_\lambda$ is injective, $\alpha_\lambda \neq \alpha_\mu$ and thus $v \otimes w \notin \ker(K)$. Combined with part (ii), this gives $\ker(K) = \mathcal{K}_G$.
\end{proof}
\end{theorem}

\begin{corollary}(Synchronization-preserving algebra)\label{cor:syncalg}

    The set $H_{\mathrm{sync}} := \{H \in \operatorname{End}_G(H) : [H,K] = 0 \}$ is a unital *-subalgebra of $B(H)$ and preserves $\mathcal{K}_G$. It is the maximal algebra of $G$-equivariant Hamiltonians preserving synchronization. 
    
    \begin{proof}
        Closure under addition, multiplication, and adjoint follows from the commutant being an algebra. If $H \in H_{\mathrm{sync}}$ and $\ket{\psi} \in K_G \subseteq \operatorname{ker}(K)$, then $K(H\ket{\psi}) = H\mathcal{K}\ket{\psi} = 0$, so $H\ket{\psi} \in \operatorname{ker}(K)$. Since $H$ also commutes with $\rho(G)$, it preserves each isotypic component, hence preserves $\mathcal{K}_G$. Maximality: any $H \notin H_{\mathrm{sync}}$ has $[H,K] \neq 0$, so there exists $\ket{\psi} \in \operatorname{ker}(K)$ with $KH\ket{\psi} \neq 0 $.
    \end{proof}
    
\end{corollary}

\begin{example} (Two qubits with $\mathbb{Z}_2$ symmetry)
    Let $G = \mathbb{Z}_2, H_A = H_B = \mathbb{C}^2$ with $\rho_A(g) = \rho_B(g) = \sigma_z$ for the nontrivial element $g$. The irreducible decomposition gives two one-dimensional representations: $V_{+} = \operatorname{span}\{\ket{0}\}$ and $V_{-} = \operatorname{span}\{\ket{1}\}$. The diagonal isotypic subspace is $\mathcal{K}_G = \operatorname{span}\{\ket{00}, \ket{11}\} = \operatorname{ker}(K)$. The algebra $H_{\mathrm{sync}}$ consists of all Hamiltonians diagonal in the computational basis, generated by $Z \otimes I, I \otimes Z$, and $Z \otimes Z$.
    
\end{example}

\begin{example} ($S_3$ acting on $\mathbb{C}^3$)
    Let $G = S_3$ act on $H_A = H_B = \mathbb{C}^3$ by permutation of basis vectors. This decomposes as $V_{triv} \oplus V_{std}$ where $V_{triv} = \operatorname{span}\{\ket{0} + \ket{1} + \ket{2}\}$ is the trivial representation and $V_{std}$ is the two-dimensional standard representation. Define $T_A = T_B$ as the projection onto $V_{triv} $(or equivalently, the element) 

    \begin{equation}
            \frac{1}{3} \sum_{g \in G} \rho(g) \in Z(\mathbb{C}[G])).
    \end{equation}

    Then $T_A$ acts as $1$ on $V_{triv}$ and $0$ on $V_{std}$, so $K = T_A \otimes I - I \otimes T_B$ annihilates exactly $\mathcal{K}_G = (V_{triv} \otimes V_{triv}) \oplus (V_{std} \otimes V_{std})$ (recalling that $T_A = T_B$ is the central idempotent projecting onto the trivial representation). The synchronization-preserving algebra consists of all $S^3$-equivariant operators commuting with $K$, which decomposes as scalar multiples of the identity on each $V_\lambda \otimes V_\lambda$. 
    
\end{example}

\begin{remark}
    The classification identifies synchronization as a structural invariant: it is determined by the alignment of irreducible components across subsystems, not by the specific norm of the observables. Any two $G$-equivariant clock observables derived from the same central element of $\mathbb{C}[G]$ yield identical synchronization subspaces. This suggests that synchronization may be viewed as a property of the representation category rather than of individual operators. 
    
\end{remark}

\section{Discussion}\label{discuss}

We have established two main results concerning the preservation and stability of synchronization subspaces defined by kernel conditions on tensor products. Theorem \ref{thm:driftbound} provides a sharp perturbative bound showing that drift from the synchronization subspace grows at most linearly in time under $\epsilon$-compatible dynamics, with the spectral gap of the synchronization operator governing fidelity decay. Theorem \ref{thm:synciso} classifies the synchronization subspace as the diagonal isotypic component under a finite group symmetry and characterizes the algebra of synchronization-preserving dynamics as the intersection of the commutant of the group action with the commutant of the synchronization operator.

These results are motivated by quantum time transfer, where distributed parties synchronize local clocks through correlated measurements on shared quantum states \cite{PhysRevApplied.20.024064, lafler2024twoway}. In that setting, the operators $T_A$ and $T_B$ model clock observables whose eigenvalues represent time labels, the kernel condition encodes perfect synchronization, and $\epsilon$-compatible dynamics capture the effect of experimental imperfections such as noise, environmental coupling, clock drift, or finite-precision control. The perturbative bounds of Section \ref{stability} provide quantitative guarantees on the timescale over which synchronization remains reliable, while the algebraic classification of Section \ref{representation} identifies the structural constraints that any synchronization-preserving protocol must satisfy. Operational aspects of such protocols have been developed in several experimental and theoretical works.

The framework generalizes naturally to multipartite systems, where synchronization subspaces arise as intersections of pairwise kernel conditions or through higher-order symmetry constraints on tensor products of multiple subsystems. The algebraic structure of synchronization-preserving dynamics in such settings, and its dependence on the topology of the network, remains to be classified. In particular, for $n$ subsystems with clock observables $\{T_1, \cdots, T_n\}$, the synchronization subspace is the intersection $\displaystyle\bigcap_{i<j} \operatorname{ker}(T_i \otimes I - I \otimes T_j)$, and one expects the structure of this intersection to depend on the spectral relationships among the $T_i$ in a way that the pairwise theory does not capture. 

The resemblance between the kernel condition and stabilizer constraints in quantum error correction \cite{Gottesman1996} raises the question of whether synchronization subspaces can be actively protected through coding schemes, potentially extending timing fidelity beyond the perturbative regime. Finally, the characterization of synchronization-preserving unitaries as elements of a commutant algebra suggests a categorical formulation in which these unitaries serve as morphisms between objects equipped with compatible observables and dynamics.

\section*{Acknowledgments}

The author thanks Dr. R. Nicholas Lanning and Mr. Aaron Marcus for detailed feedback on earlier drafts of this paper and helpful discussions.

\section*{Declarations}

\textbf{Funding.} The author did not receive support from any organization for the submitted work.

\textbf{Competing interests.} The author has no relevant financial or non-financial interests to disclose.

\textbf{Data availability.} Data sharing is not applicable to this article as no datasets were generated or analyzed.

%\begin{appendices}

%\section{Section title of first appendix}\label{secA1}




%\end{appendices}


\bibliography{sn-bibliography}% common bib file

\end{document}